\documentclass{aastex61}
\usepackage{graphicx}
\usepackage{longtable}
\usepackage{amsmath}
\usepackage[utf8]{inputenc}
\usepackage{amssymb}
\usepackage{subfigure}
\usepackage{hyperref}
\usepackage{dcolumn}
\usepackage{bm}
\usepackage{color}
\usepackage{cleveref}

\newcommand{\mpl}{{M_{\rm {pl}}}}

\newcommand{\dd}{\, {\rm d}}
\newcommand{\gsim}{\;\mbox{\raisebox{-0.5ex}{$\stackrel{>}{\scriptstyle{\sim}}$}
}\;}
\newcommand{\lsim}{\;\mbox{\raisebox{-0.5ex}{$\stackrel{<}{\scriptstyle{\sim}}$}
}\;}

\newcommand{\pn}{\Phi_{\rm N}}

\newcommand*\colvec[3][]{\begin{pmatrix}\ifx\relax#1\relax\else#1\\\fi#2\\#3\end{pmatrix}}

\newcommand{\oo}{\mathcal{O}}

\newcommand{\abh}{a_{\rm BH}}
\newcommand{\rv}{r_{\rm v}}

\def\eea{\end{eqnarray}}
\def\bea{\begin{eqnarray}}
\allowdisplaybreaks
\usepackage{enumitem}
\setenumerate[1]{label=(\arabic*)}

\begin{document}
\title{Tests of Gravity Theories Using Supermassive Black Holes} 
\author{Jeremy Sakstein}
\email{sakstein@physics.upenn.edu}
\affiliation{Department of Physics and Astronomy, Center for Particle Cosmology,
University of Pennsylvania, 209 S. 33rd St., Philadelphia, PA 19104, USA}
\author{Bhuvnesh Jain}
\email{bjain@physics.upenn.edu }
\affiliation{Department of Physics and Astronomy, Center for Particle Cosmology,
University of Pennsylvania, 209 S. 33rd St., Philadelphia, PA 19104, USA}
\author{Jeremy S. Heyl}
\email{heyl@phas.ubc.ca}
\affiliation{Department of Physics and Astronomy, University of British Columbia, 6224 Agricultural Road, Vancouver, BC V6T 1Z1, Canada}
\author{Lam Hui}
\email{lhui@astro.columbia.edu }
\affiliation{Department of Physics, Center for Theoretical Physics,
Columbia University, New York, NY 10027, USA}

\begin{abstract}
Scalar-tensor theories of gravity generally violate the strong equivalence principle, namely compact objects have a suppressed coupling to the scalar force, causing them to fall slower. A black hole is the extreme example where such a coupling vanishes, i.e. black holes have no scalar hair. We explore observational scenarios for detecting strong equivalence principle violation, focusing on galileon gravity as an example. For galaxies in-falling towards galaxy clusters, the supermassive black hole can be offset from the galaxy center away from the direction of the cluster. Well resolved images of galaxies around nearby clusters can therefore be used to identify the displaced black hole via the star cluster bound to it. We show that this signal is accessible with imaging surveys, both ongoing ones such as the Dark Energy Survey, and future ground and space based surveys. Already, the observation of the central black hole in M~87 places new constraints on the galileon parameters, which we present here. $\mathcal{O}(1)$ matter couplings are disfavored for a large region of the parameter space. We also find a novel phenomenon whereby the black hole can escape the galaxy completely in less than one billion years. 
\end{abstract}

\section{Introduction}

The study of infrared modifications of general relativity (GR) has undergone a renaissance in the last decade, driven partly by the dark energy mystery and partly by recent developments in the construction of healthy higher-derivative scalar-tensor theories (see \citet{Clifton:2011jh,Joyce:2014kja,Koyama:2015vza,Berti:2015itd,Bull:2015stt,Joyce:2016vqv,Burrage:2016bwy} for recent reviews), which have opened up a new realm of possibilities for driving the acceleration of the cosmic expansion. On small scales, consistency with tests of GR is achieved by utilising screening mechanisms, which use non-linear effects to hide the modifications of gravity in the solar system. One particularly interesting and well-studied paragon for these theories is the galileon \citep{Nicolis:2008in}, which self-accelerates cosmologically but screens locally using the Vainshtein mechanism \citep{Vainshtein:1972sx}. On large scales, the galileon can be tested by the modified rate of structure growth \citep{Barreira:2012kk,Barreira:2013xea}; on small scales, the Vainshtein mechanism renders its predictions largely indistinguishable from those of GR.

One interesting exception is that of \citet{Hui:2012jb}, who have pointed out that the no hair theorem \citep{Hui:2012qt} implies that black holes do not couple to the galileon, and therefore there is a violation of the strong equivalence principle (SEP) whereby they fall at different rates to non-relativistic matter in external fields. In particular, if part of a galaxy's motion is due to an external galileon field then the supermassive black hole (SMBH) that lies at its center does not feel this, and therefore the SMBH lags behind as the galaxy moves. \citet{Hui:2012jb} have argued that, for constant density cores, the restoring force due to the dark matter and baryons in the galaxy will eventually compensate for the lack of the galileon force, giving rise to an offset in the opposite direction to the galaxy's acceleration. \cite{Hui:2012jb} considered galaxies moving in the cosmic field, and predicted a small offset, which is difficult to look for observationally. 

In this letter, we look for novel scenarios for testing galileon gravity using the predicted SEP violation that are more amenable to observational testing. We consider two new effects:
\begin{enumerate}
\item {\bf Galaxy clusters}: Satellite galaxies in-falling into galaxy clusters pass through a partially unscreened regime where the galileon field generated by the cluster
pulls on the galaxies but not their resident black holes. This can result in an offset which is accessible to ongoing and planned imaging surveys. In this work we use the analysis of M~87 presented in \citep{Asvathaman:2015nna} to place new constraints.
\item {\bf Escaping black holes}: Realistic dark matter haloes whose density falls with distance from the center have a maximum restoring force. Galileon forces larger than this drive the SMBH to escape the galaxy in an observable time-scale (less than one billion years).
\end{enumerate}

\section{Galileon Gravity Theories}

Galileons are scalar-tensor extensions of GR that are invariant under the Galilean transformation $\phi(x^\mu)\rightarrow\phi(x^\mu)+b_\mu x^\mu+c$  \citep{Nicolis:2008in}. This symmetry restricts the form of their action to four unique terms (in four dimensions), the so-called quadratic, cubic, quartic, and quintic galileons \cite{Nicolis:2008in}, and ensures that the equations of motion are second-order so that the Ostrogradski ghost instability is absent. In what follows, it will be sufficient to restrict our attention to the simplest model, the cubic galileon, although we note that the results presented here apply equally to all galileon theories including their curved space generalisations \citep{Deffayet:2009wt,Deffayet:2011gz,Gleyzes:2014dya,Gleyzes:2014qga}. In more general cases, there will be additional parameters and the tests we discuss later will constrain combinations of these.
The equation of motion for the cubic galileon field sourced by a static non-relativistic object is
\begin{equation}\label{eq:eom}
\nabla^2\phi+\frac{r_c^2}{3}\left[\left(\nabla^2\phi\right)^2-\nabla_i\nabla_j\phi\nabla^i\nabla^j\phi\right]={8\pi \alpha G}\rho,
\end{equation}
where $i,\,j=1,\,2,\,3$. 
There are two constant parameters\footnote{If one considers the cosmological evolution then these become weak functions of redshift but this dependence is sub-leading provided that one considers the most general galileon theories (as one should), which include an $\oo(1)$ constant contribution to $\alpha$ coming from a Brans-Dickie-like coupling. The tests we will discuss here are only sensitive to the sum of the constant and time-dependent pieces, and we take them to be dominated by the constant piece for simplicity. }: a dimensionless coupling constant $\alpha\sim\oo(1)$, which parameterizes the non-minimal coupling to matter, and the \emph{crossover scale} $r_c$, which determines the size of the non-linear galileon self-interactions. The coupling scale $\Lambda_3=(6\mpl/r_c^2)^{1/3}$ is often used instead of $r_c$. For a point-mass $M$, one can define the \emph{Vainshtein radius} $\rv^3\equiv\frac{4}{3}\alpha GMr_c^2$, which is the scale where the non-linear galileon terms become important. The total (Newtonian plus galileon) force profile inside the Vainshtein radius is
\begin{equation}\label{eq:vain}
\frac{\dd\Phi}{\dd r}=\frac{GM}{r^2}\left[1+2\alpha^2\left(\frac{r}{\rv}\right)^{\frac{3}{2}}\right]
\end{equation}
so that deviations from GR are suppressed by a factor $(r/\rv)^{3/2}$. Outside the Vainshtein radius the galileon force becomes unsuppressed,
leading to a total force profile that matches the Newtonian one, except for an overall enhancement by a factor of $(1+2\alpha^2)$. A solar mass object has $\rv\sim\mathcal{O}(100 \textrm{ pc})$ \citep{Khoury:2013tda} and so the solar system lies well within the Sun's Vainshtein radius. For extended objects, such as the dark matter haloes we discuss in this work, the transition
between the screened and unscreened regimes is a rather gradual one (see \cite{Schmidt:2010jr} and fig. \ref{fig:forcerat} below). This is one reason why the outskirts of galaxy clusters are a prime testing ground for galileon theories.

In this theory, the equation of motion for an object of mass $m$ moving in an external Newtonian and scalar potential, $\pn^{\rm ext}$ and $\phi^{\rm ext}$ respectively, sourced by some other object is $m\ddot{\vec{x}}=-m\nabla\pn^{\rm ext} - \alpha Q\nabla\phi^{\rm ext}$, where $Q$ is the scalar charge that characterises the strength of the object's coupling to the galileon field. One can show \citep{Hui:2009kc} that $Q=\int\dd^3x\, T^0_0$ i.e. the scalar charge is equal to the baryonic mass\footnote{We use the term ``baryons" as a proxy for anything that contributes to the matter energy-momentum tensor (as opposed to the pseudo-energy-momentum tensor which includes the contribution from gravitational binding energy). This could include actual baryons as well as dark matter.}
of the object. This implies that $Q=m$ for non-relativistic objects but that $Q<m$ for compact objects, whose mass receives a significant contribution from the gravitational binding energy. A black hole is the extreme example where $Q=0$. \citet{Hui:2012qt} showed that a black hole cannot source galileon hair, and therefore also does not couple to a galileon external field. Since our primary interest concerns a black hole residing in a galaxy, one might worry that the Vainshtein mechanism is enough to greatly diminish the galileon force on the stars (and dark matter) of the host galaxy, leading to essentially no difference between the falling motion of the stars and that of the black hole.
A key feature of the galileon comes to the rescue: the galileon equation of motion is invariant under the galileon symmetry and so, given any solution, one can always add a component with a constant field gradient to obtain a second solution. For instance, for a galaxy located within a cluster, the galileon field sourced by the cluster behaves as a constant-gradient field, since its scale/wavelength is much longer than the size of the galaxy \citep{Hui:2009kc}. The resident stars and dark matter of the galaxy thus respond to this cluster-sourced galileon field, while the black hole does not, setting up an astronomical version of the E\"otv\"os experiment. Note that in this set-up, it is important to account for the effect of Vainshtein mechanism on the cluster-sourced galileon field itself.

\begin{figure}[ht]\centering
{\includegraphics[width=0.45\textwidth]{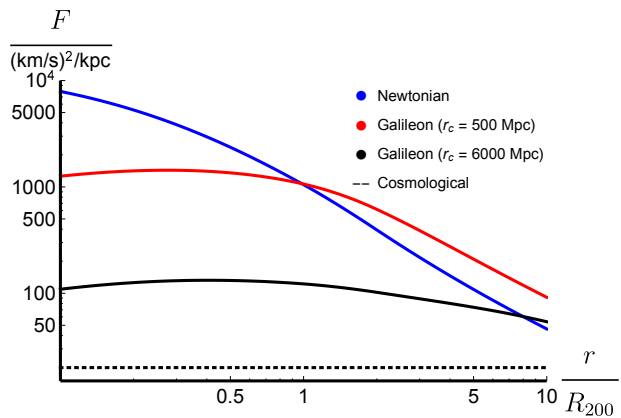}}
\caption{The Newtonian, galileon, and cosmological forces for a cluster with mass $M_{200}=10^{15} M_\odot$. The galileon force falls below the Newtonian force inside the Vainshtein radius as expected, but it remains flat even well inside the cluster. For black hole offsets, the true amplitude of the galileon force matters (unlike other tests which rely on its amplitude relative to the Newtonian force). }\label{fig:forcerat}
\end{figure}

\section{Galaxy Clusters}

A cluster carries sufficient mass that, despite some Vainshtein suppression, the galileon force sourced by it is large
enough to give interesting observational effects at distances $D\gsim 0.1R_{200}$ \citep{Schmidt:2010jr}.
As an example, consider a model of the Virgo cluster with mass $M_{200}=10^{15}M_\odot$. We model the mass distribution of the cluster with a Navarro-Frenk-White (NFW) profile \citep{Navarro:1995iw} (with concentration $c=5$) inside $R_{200}$; outside it we model the ``2-halo'' mass distribution using the fits to N-body simulations of \citet{Diemer:2014xya}. We define $R_{200}$ using the critical density, which corresponds to $R_{200c}$ in \citet{Diemer:2014xya}; we take $H_0=72$ km/s/Mpc. Note that we consider haloes at $z=0$ since the tests described presently are intended to be applied to low redshift ($z<0.05$) clusters. Any potential time-evolution of $\alpha$ over this range will be negligible, although we note that more detailed modelling may be required if a sample of useful clusters at redshifts $z\gg0.1$ were to be found. Previous studies have consistently found that the properties of halos are largely unmodified in galileon cosmologies \citep{Barreira:2014zza,Barreira:2015fpa}, especially inside the Vainshtein radius. Outside, at distances larger than $\sim 2$ Mpc, more detailed modelling may be required, although M~87, which we consider below, lies well within the Vainshtein radius of the Virgo cluster so this is not a caveat to the results we obtain here. 

The Newtonian, galileon, and cosmological root mean square (RMS) galileon forces (the scenario considered by \citet{Hui:2012qt}) are plotted in Fig.~\ref{fig:forcerat} for models with $r_c=500$ and $6000$ Mpc\footnote{The latter corresponds to self-accelerating cosmological galileons while the former is often used as a paradigm for Vainshtein screening \citep{Schmidt:2010jr}. These models are not ruled out by local observational constraints \citep{Dvali:2002vf,Khoury:2013tda}. The constraints of \citet{Schmidt:2009sv,Barreira:2014jha} coming from cosmological probes do not directly apply to our model since they are derived for a sub-class of models where the constant contribution to $\alpha$ is absent. Future efforts to constrain more general galileon models, in particular those with a linear coupling to matter, will be able to use the results we obtain below as a consistency check.}. The galileon force was calculated by solving the field equations for the cubic galileon exactly using spherical symmetry. One can see that there is a large galileon force for $r\gsim 0.1R_{200}\sim 0.3$ Mpc. Satellite galaxies are typically in-falling towards the cluster within $\lsim 4R_{200}\sim13$ Mpc \citep{Diemer:2014xya}. These galaxies would feel a large galileon force that their central SMBHs would not, thus allowing the effects of the SEP violation to manifest. Note that the galileon force is suppressed relative to the Newtonian force when $r\lsim R_{200}$ but it is the amplitude of the galileon force that is important for SEP violations (not this ratio) and so large effects are still expected at distances above $ 0.1 R_{200}$. The cluster forces are at least an order of magnitude larger than RMS cosmological force. One therefore expects offsets of $\oo(\textrm{ kpc})$, which are accessible to imaging surveys of nearby galaxy clusters. 

\begin{figure}[h]\centering
{\includegraphics[width=0.45\textwidth]{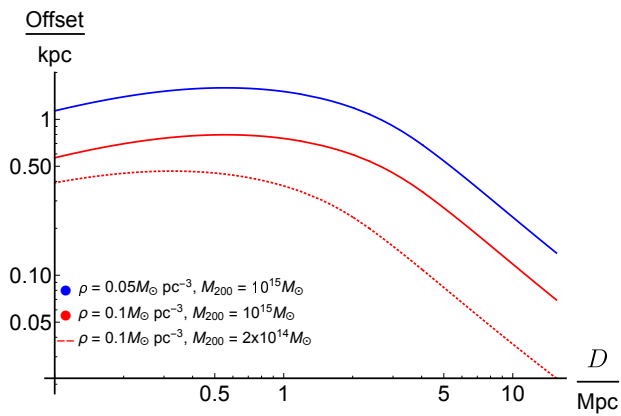}}
\caption{The offset of the SMBH for a cored galaxy in-falling towards a cluster. The central densities (of the in-falling satellites) and the cluster mass ($M_{200}$) are indicated in the figure. In all cases we have set $c=5$, $r_c=500$ Mpc, and $\alpha=1$; we used the same modelling techniques as in fig. \ref{fig:forcerat}. The curves show that the outskirts of clusters are promising for this test as the offsets are large, and most of the in-falling galaxies have not had interactions with other galaxies that may perturb the SMBH.  }\label{fig:offset_const}
\end{figure}

The dynamics of this situation are easily exemplified by considering a galaxy where the central density $\rho_0$ is approximately constant. In this case, the black hole is subject to the galileon acceleration $\abh$ away from the center and a linear restoring force given by $4/3\pi G r\rho_0$, in which case 
the equation of motion for the black hole (in the rest frame of the galaxy) is
\begin{equation}
\ddot{r}= - \frac{4\pi}{3}G\rho_0 r + \abh
\end{equation}
so that the black hole undergoes oscillations about the equilibrium point $\bar{r}=3\abh/4\pi G\rho_0$ with a time-period $T_0\sim (G\rho_0)^{-1/2}$. In practice, the galileon force turns on slowly over a time $\sim \dot{a}_{\rm BH}/\abh\sim v/d\sim \mathcal{O}(10^{10}\textrm{ years})$ (where $v\sim100$ km/s is the typical in-fall velocity and $d$ is the distance from the cluster center) whereas the oscillation period is $\mathcal{O}(10^7\textrm{ years})$ so that the amplitude of these oscillations is small and the black hole tracks the equilibrium position adiabatically. One then expects the black hole to be offset from the center by a distance
\begin{equation}\label{eq:offsetcore}
\bar{r}=1 \textrm{ kpc } \left(\frac{\abh}{2000(\textrm{km/s})^2/\textrm{kpc}}\right)\left(\frac{\rho_0}{0.1 M_\odot\textrm{pc}^{-3}}\right)^{-1}.
\end{equation}
Clearly the size of the offset depends on the central density $\rho_0$ and the external force; we have used a fiducial external force that is typical for the outskirts of a galaxy cluster similar to the one modelled in fig. \ref{fig:forcerat}. As the galaxy in-falls, the galileon force increases and so does the offset of the black hole (up to around $R_{200}$); the direction of the offset would be in the opposite direction to the galaxy's direction of acceleration i.e. away from the center of the cluster. 

An example of this (taking $\alpha=1$) is shown in Fig.~\ref{fig:offset_const}; one can see that observable offsets of $\mathcal{O}(\textrm{kpc})$ can develop at a variety of distances from the cluster's centre. They are at least an order of magnitude larger than one would expect for galaxies moving in the cosmological field, and so this scenario is well suited to testing galileon gravity. As is evident form the figure, as the galaxy in-falls, the offset becomes larger due to an increased galileon acceleration. This reaches a maximum and begins to fall off as the galaxy approaches the cluster's center, where the galileon modifications are more screened.

\begin{figure}[ht]\centering
{\includegraphics[width=0.45\textwidth]{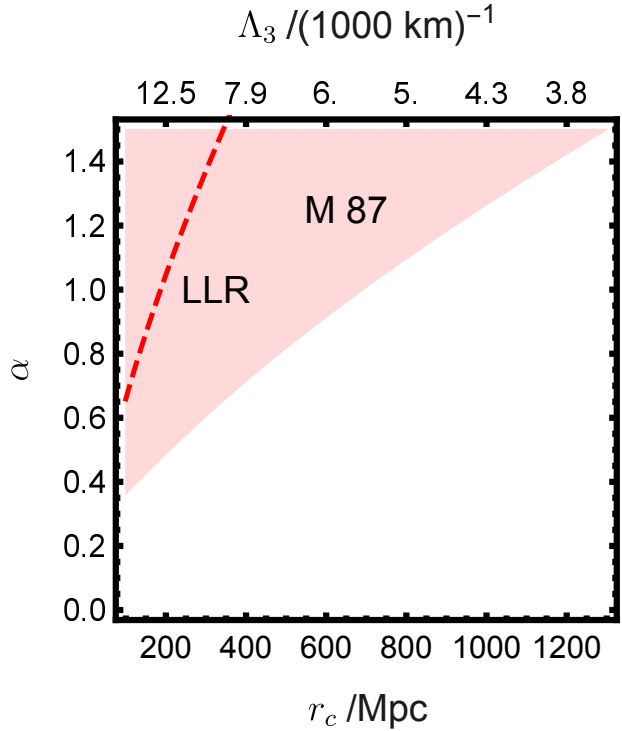}}
\caption{The white area depicts the allowed parameter space of the cubic galileon model. We obtain these constraints from the result of \citep{Asvathaman:2015nna} that the central black hole of M~87 is displaced by less than 0.03~arcseconds from the centroid of the galaxy, yielding a constraint of $F < 10^3~ (\textrm{km/s})^2/\textrm{kpc}$. The dashed red line shows the previous bounds from lunar laser ranging measurements \citep{Dvali:2002vf}. Note that self-accelerating galileon models typically have $r_c>1000$ Mpc. }\label{fig:cubic_constraints}
\end{figure}

\cite{Asvathaman:2015nna} considered the effect of the sub-clump of the Virgo cluster centered on M~84 and M~86 on the position of the black hole within M~87. Using the same assumptions as \citet{Asvathaman:2015nna}, in particular the dynamical model of \citet{0004-637X-770-2-86} and the photometry of \citet{2006ApJS..164..334F}, we find that the strength of the galileon field in the plane of the sky is less than about $700~ (\mathrm{km/s})^2/\mathrm{kpc}$ (or $1000~(\textrm{km/s})^2/\textrm{kpc}$ in three dimensions). We find the SMBH to be located at the centre of the light to within 0.03 arcseconds (0.03 is the upper limit at 1$\sigma$). We are only sensitive to the motion of the SMBH across the sky because we are measuring its position on the sky relative to the center of the galaxy; consequently, only the component of the galileon field in the plane of the sky is measurable. The distribution of light and mass-to-light ratio in the context of a Hernquist model for the central region of the galaxy was used to obtain an estimate of the restoring force. 

Combining this with a model for the Virgo cluster as depicted in Fig.~\ref{fig:forcerat} i.e. NFW + 2 halo modelling with $c=5$ and $M_{200}=10^{15}M_\odot$ \footnote{masses in the range $M_{200}=1$--$2\times10^{15}M_\odot$ have been reported \citep{Fouque:2001qc,Peirani:2005ti}, which is less uncertain than effects of modelling the complex 3D structure \citep{Mei:2007xs} with an NFW profile so we adopt the conservative value of $10^{15}M_\odot$.} we obtain constraints on the values of $\alpha$ and $r_c$ as depicted in Fig.~\ref{fig:cubic_constraints}. These were found by scanning the parameter space to find the region where the galileon force at $R_{200}$ is smaller than $1000~(\textrm{km/s})^2/\textrm{kpc}$. The constraints therefore assume that M~87 is located within $R_{\rm 200}$ of the Virgo cluster, but
is otherwise insensitive to its precise location because the galileon force is fairly constant for $r < R_{\rm 200}$. We have checked that varying the concentration over the range $c=5$--$10$ produces a minimal change in the constraints. Finer constraints could be obtained by measuring the displacements of many SMBHs relative to the centers of their galaxies and the local density field and averaging over an ensemble. Such an analysis could probe theories with $r_c>1500$ Mpc, however, the observations of M~87 already yields tighter constraints on this model than solar system measurements. There are several caveats to our analysis, including the use of spherical symmetry in the predictions \citep{Asvathaman:2015nna}. We leave a detailed analysis of the observational strategy and implementation for future work.
%

The constant density approximation works well at small distances but the offsets predicted in Fig. \ref{fig:offset_const} lie well in the regime where the fall-off of the density profile with distance becomes important. The density near the center of galaxies tends to fall off with some power $\rho\propto r^{-\beta}$ with $\beta<1$ \citep{Trujillo:2004ga,2006ApJS..164..334F} that steepens at larger radii. This results in a rising restoring force that reaches some maximum and begins to decrease when the profile steepens\footnote{Profiles with $\beta\ge1$ result in maximum or divergent forces at the centre but, in practice, the SMBH will drag a disc of stars with it so that the restoring force vanishes at the centre \citep{Asvathaman:2015nna}. The situation described above is therefore generic. }. Small galileon forces, such as those generated by the cosmological large scale structure, are not sufficient to overcome the maximum restoring force and hence lead to an offset. Larger forces, such as those that can be generated by clusters can overcome the maximum restoring force and can therefore cause the black hole to escape the galaxy completely. Even for the case that the restoring force is larger at the center, a black hole that is displaced by other mechanisms can be driven out of the galaxy by the galileon force. Such a situation might arise as a result of a merger event or by recoil from gravitational wave ``kicks".

In practice, if the in-falling galaxy is initially at a position where the galileon force is smaller than the restoring force then the black hole will be offset but, as the galaxy moves closer to the cluster, the galileon force will increase and may exceed the restoring force so that there is nothing halting the SMBH's motion. One can estimate the time-scale for the black hole to escape the galaxy entirely by neglecting the restoring force. Galaxies falling into clusters experience an increasing galileon force whilst the restoring force remains the same and so this quickly becomes a good approximation. Treating the galileon acceleration as constant, the time-scale for the black hole to move a distance $R$ is
\begin{equation}
T={10^7 \textrm{ yr}}\left(\frac{R}{ \textrm{kpc}}\right)^{\frac{1}{2}}\left(\frac{\abh}{2000 (\textrm{km/s})^2/\textrm{kpc}}\right)^{-\frac{1}{2}}.
\end{equation}
The black hole can then escape the galaxy in less than a billion years. Note that the typical velocity of the black hole is $\oo(\textrm{km/s})$, far less than the typical velocity dispersion of the galaxy so that dynamical friction can safely be ignored \citep{2008gady.book.....B}. Precisely which galaxies would allow escape depends on the environment and the galaxy central density. We leave a systematic investigation for the future.

\section{Observational Tests}

The novel features identified above present new avenues for testing galileon gravity and constraining the parameters $\alpha$ and $r_c$ (recall that $\alpha$ parameterises the strength of the galileon coupling to matter and $r_c$ parameterises the galileon's self-interactions).  

{\it Displaced SMBHs in nearby galaxies}: The gravitational acceleration is several times larger at the outskirts of galaxy clusters than in the field, and is directed towards the cluster center. The predicted offset is detectable via the star cluster bound to the SMBH, which would be offset from the centroid of the stellar light at the center of the galaxy. The correlation with the expected displacement direction from modified gravity, i.e. the opposite direction to the galaxy cluster, is critical in such an exercise as it narrows the search zone and can distinguish it from other sources of fluctuations. 

The two key observational parameters are the size of the offset relative to the point spread function (PSF), which determines whether it is resolved, and the change in flux due to the star cluster bound to the SMBH. At $z\simeq 0.05$, an offset of 1 kpc corresponds to 1 arcsecond on the sky, which is larger than the (full width half maximum) PSF of the best ground based telescopes. Out to this distance, hundreds of clusters can be found for which the offsets in in-falling galaxies are resolved. A large fraction of these are already imaged by the SDSS and DES optical surveys and by X-Ray surveys. By identifying suitable galaxies at the outskirts of each cluster, a large sample of galaxies can be assembled. These galaxies could be followed up with high resolution imaging, ideally from the Hubble Space Telescope, and multi-wavelength observations. The sample size may be essential in handling the second observational challenge: detecting the star cluster given the typical flux variations across a galaxy, and other sources of error. For the central parts of elliptical galaxies, these are at the percent level (e.g. \citet{Bernardi:2017qfv}), which in many cases is smaller than the displaced SMBH's star cluster (e.g. \citet{Asvathaman:2015nna}'s study of M87). Another challenge discussed by \citet{Asvathaman:2015nna} is centroiding the galaxy light using the outer isophotes. Finally, we note that the predicted offset is sensitive to the central density profile of the host galaxy, which is challenging to determine. A range of offset values must therefore be considered, and for lower values only the most nearby clusters may be suitable for our test.

{\it Perturbations to the galaxy light profile}: while we have not investigated morphological features in this study, a SMBH that is displaced or drifting through the stellar disk will produce characteristic distortions that can be measured by analyzing the images of a large sample of galaxies. Even for galaxies above $z\sim 0.1$, where the offset is no larger than the PSF, the model fitting approach described above can be attempted. Weak lensing studies that measure the low order moments of the surface brightness of galaxies may also be well suited to extracting the skewness or ``flexion'' in the light distribution, and correlating it with the direction of the external force vector on the galaxy. 

{\it Missing SMBHs}: for larger galileon forces, or if initially displaced, the SMBH would leave the visible galaxy in less than a billion years, the timescale over which galaxies move with coherent velocities. Hence, a fraction of galaxies would not have central SMBHs in such a scenario. During galaxy mergers, SMBHs are displaced from their center and then occupy steeper parts of the density profile until they lose energy to dynamical friction and sink to the center. The timescales may be sufficient for the modified gravity effect to act on the black hole. The observational finding that the overwhelming majority of galaxies above a certain mass have central SMBHs may provide new limits on $\alpha$ and $r_c$ for typical values of the density profile. 

Note that for many of these observable tests, we can estimate the direction of the effect from observations of the galaxies surrounding the cluster of interest by estimating the direction of the local gravity acceleration vector. The direction vector is more reliable for galaxies in-falling into galaxy clusters. This is crucial for distinguishing modified gravity effects from other astrophysical processes such as acceleration due to asymmetric jets, recoil from gravitational   wave emission, Brownian motion, gravitational slingshot due to mergers, and perturbations due to massive objects such as globular clusters. All of these can displace the central black hole for varying periods of time, but in a direction uncorrelated with the galaxy's acceleration. These processes are important to study even in the absence of modified gravity and may shed light on the dynamics of galaxy mergers \citep{Merritt:2004gc}. 

\section{Conclusions}

To summarize, we have identified novel methods for testing gravity using the equivalence principle violations for black holes first noted by \citet{Hui:2012jb}. Galaxies in-falling into massive clusters may host SMBHs that are offset from the center by $\mathcal{O}(\textrm{kpc})$ and may even be absent altogether. The offset is at least an order of magnitude larger than for galaxies moving in the cosmological field, and is in the opposite direction to the galaxy's acceleration, which can help distinguish this effect from other astrophysical displacement mechanisms. We have discussed how a sample of nearby galaxy clusters can be used to obtain bounds on the model parameters $\alpha$ and $r_c$, which would have strong implications for models of modified gravity that try to explain cosmic acceleration. Indeed, using observations of the central black hole in M~87 we have been able to place new constraints on the parameter space that push into the interesting region for self-accelerating models. 

{\it Acknowledgements: } We are grateful to 
Eric Baxter, Mariangela Bernardi, Gary Bernstein, Chihway Chang, Neal Dalal, Marla Geha, Jenny Green, Mike Jarvis, Gordon Richards, Ed Moran, Nadia Zakamska, Justin Alsing and colleagues at the Center for Computational Astrophysics. JS is supported by funds provided to the Center for Particle Cosmology by the University of Pennsylvania. BJ is supported in part by the US Department of Energy grant DE-SC0007901. 
JH is supported by a Discovery Grant from the Natural Sciences and Engineering Research Council of Canada. LH is supported in part by the NASA grant NXX16AB27G and the DOE grant DE-SC0011941.


\end{document}